\documentclass[<1p>]{elsarticle} 
\usepackage{amsmath, amsthm, amssymb, amsthm}
\usepackage{subcaption}
\usepackage[flushleft]{threeparttable}
\usepackage{tabu,multirow}
\usepackage{color}
\usepackage{etoolbox}
\usepackage{float}
\usepackage{booktabs}
\usepackage{etoolbox}
\let\today\relax
\usepackage{etoolbox}
\usepackage{etoolbox}
\usepackage{siunitx}
\patchcmd{\MaketitleBox}{\footnotesize\itshape\elsaddress\par\vskip36pt}{\footnotesize\itshape\elsaddress\par\parbox[b][36pt]{\linewidth}{\vfill\hfill\textnormal{\today}\hfill\null\vfill}}{}{}%
\patchcmd{\pprintMaketitle}{\footnotesize\itshape\elsaddress\par\vskip36pt}{\footnotesize\itshape\elsaddress\par\parbox[b][36pt]{\linewidth}{\vfill\hfill\textnormal{\today}\hfill\null\vfill}}{}{}%
\makeatletter
\def\ps@pprintTitle{%
    \let\@oddhead\@empty
    \let\@evenhead\@empty
    \def\@oddfoot{\hfill\footnotesize\itshape 8 January 2024}%
    \let\@evenfoot\@oddfoot
}
\usepackage{hyperref}
\makeatother

\begin{document}
\begin{frontmatter}
\begin{abstract}
This paper investigates the performance of the Generalized Covariance estimator ($GCov$) in estimating and identifying mixed causal and noncausal models. The $GCov$ estimator is a semi-parametric method that minimizes an objective function without making any assumptions about the error distribution and is based on nonlinear autocovariances to identify the causal and noncausal orders. When the number and type of nonlinear autocovariances included in the objective function of a $GCov$ estimator is insufficient/inadequate,  or the error density is too close to the Gaussian, identification issues can arise. These issues result in local minima in the objective function, which correspond to parameter values associated with incorrect causal and noncausal orders. Then, depending on the starting point and the optimization algorithm employed, the algorithm can converge to a local minimum. The paper proposes the use of the Simulated Annealing (SA) optimization algorithm as an alternative to conventional numerical optimization methods. The results demonstrate that SA performs well when applied to mixed causal and noncausal models, successfully eliminating the effects of local minima. The proposed approach is illustrated by an empirical application involving a bivariate commodity price series.
\end{abstract}
\title{Optimization of the Generalized Covariance Estimator in Noncausal Processes}
\date{\today}
\author[a]{Gianluca Cubadda}
\author[a]{Francesco Giancaterini}
\author[b]{Alain Hecq}
\author[c]{Joann Jasiak}
\address[a]{Università di Roma ”Tor Vergata”, Italy}
\address[b]{Maastricht University, The Netherlands}
\address[c]{York University, Canada}
\renewcommand{\thefootnote}{\textdagger} 
\end{frontmatter}

\section{\textbf{Introduction}}
In recent years, there has been a growing interest in employing mixed causal and noncausal processes for the analysis of economic and financial time series (\cite{lanne2013noncausal}, \cite{hecq2016identification}, \cite{gourieroux2022nonlinear}, \cite{gourieroux2022nonlinear}, \cite{rygh2022causal}, and \cite{cavaliere2020bootstrapping}). These models have gained popularity due to their ability to incorporate both causal and noncausal components, enabling them to accurately capture the intricate nonlinear dynamics feature of economic and financial phenomena. In economics, the integration of both causal and noncausal components is particularly valuable when it comes to modeling expectations. Unlike the traditional autoregressive models, the mixed models offer insights into how economic variables are influenced by the expectations and mechanisms underlying the formation of these expectations (see \cite{lanne2011noncausal}, \cite{lanne2013noncausal}). In the financial domain, these models excel in capturing crucial phenomena characterized by nonlinear dynamics, including local trends, commonly referred to as speculative bubbles. Speculative bubbles are explosive financial patterns that frequently emerge in highly volatile markets, like the cryptocurrency and commodity markets. These bubbles represent periods of excessive prices (rates), driven more by market psychology and speculation than underlying fundamentals. During such episodes, asset prices surge to unsustainable levels, often followed by a sudden and sharp decline, leading to adverse economic outcomes. Hence, the identification and analysis of these speculative bubbles are of crucial importance due to their potential to disrupt economic stability and resource allocation (see \cite{gourieroux2017noncausal}, \cite{fries2019mixed}, and \cite{hecq2021forecasting}).\\
\indent The estimation techniques available for mixed causal and noncausal processes fall into two main categories: parametric and semi-parametric estimators. Parametric methods, such as the Maximum Likelihood estimator (\cite{lanne2013noncausal} and \cite{davis2020noncausal}), yield efficient estimates when the error distribution is correctly specified but become inefficient or even inconsistent when this assumption is violated. In contrast, semi-parametric methods offer the advantage of robustness to specification errors and do not require distributional assumption on the errors of the model (\cite{gourieroux2017noncausal}, \cite{gourieroux2022generalized}, and \cite{hecq2022spectral}). As a consequence, the benefit of employing a semi-parametric estimator is evident. Currently, the only time domain semi-parametric estimator available for estimating mixed causal and noncausal models is the Generalized Covariance estimator ($GCov$). It was initially introduced by \cite{gourieroux2017noncausal} in the context of noncausal models and subsequently extended to a semi-parametrically efficient estimator in \cite{gourieroux2022generalized}. The $GCov$ is characterized by an objective function based on the autocovariances of linear and nonlinear functions of independent and identically distributed ($i.i.d.$) errors.\\
\indent This paper aims to address potential challenges associated with local minimum issues in the objective function of the $GCov$ estimator when applied to mixed causal and noncausal models. In particular, we show that these challenges may arise from difficulties in distinguishing between causal and noncausal dynamics, often linked to factors such as an insufficient number of nonlinear autocovariances, inappropriate nonlinear transformations, or a close to Gaussian error density. As a result, our findings indicate that traditional optimization algorithms, like the Broyden Fletcher Goldfarb Shanno (BFGS) algorithm, struggle to converge to the global minimum in this context. This difficulty results in the generation of inaccurate results, underscoring the limitations of conventional optimization methods when applying the $GCov$ to processes involving a noncausal component.\\
\indent To address the issues related to local minima and avoid the misestimation of our processes, we propose combining the $GCov$ estimator with the Simulated Annealing (SA) optimization algorithm. SA is a powerful metaheuristic method designed specifically to capture the global minimum when the objective function contains numerous local minima. Initially proposed by \cite{kirkpatrick1983optimization}, SA draws inspiration from the annealing process of solids to tackle optimization problems. Over the years, SA has demonstrated remarkable success in solving complex optimization problems in various fields, including computer (VLSI) design, image processing, molecular physics, and chemistry (see, for instance, \cite{wong2012simulated}, \cite{carnevali1987image}, \cite{jones1991molecular}, and \cite{pannetier1990prediction}). By integrating SA with the $GCov$ estimator, this paper presents a significant improvement in the performance of the $GCov$ estimator applied to the mixed causal and noncausal models. \\
\indent It is important to note that this paper exclusively focuses on the estimation of causal and noncausal parameters. We do not consider inference on estimated parameters neither the significance of $GCov$ Portmanteau type tests (see \cite{gourieroux2022generalized}, and \cite{jasiakneyazi}, respectively). Indeed these latter important points are dependent on the correct identification and specification of those models. Also, this paper primarily studies the application of the $GCov$ estimator for multivariate models. We also present some results for univariate processes with the purpose of illustrating the graphical behavior of the objective function estimator when expressed as a function of a single parameter. In this context, alternative strategies may be employed to achieve a successful convergence of $GCov$. For example, a grid search strategy over the initial values to then select the estimation that yields a lower objective function value (see \cite{bec2020mixed} for a grid search approach when applied to the parametric framework). However, applying this alternative methodology in the multivariate framework can be challenging due to the large dimensions of the grid arrangement.\\
\indent The rest of the paper is organized as follows. Section 2 discusses mixed causal and noncausal models and introduces the $GCov$ estimator. Section 3 shows that its objective function may exhibit local minima under some conditions. Section 4 suggests the use of SA to overcome the issue of local minima and optimize the choice of initial values. Section 5 investigates a bivariate commodity price series. Section 6 concludes.

\section{\textbf{Estimation of mixed causal and noncausal processes by the GCov estimator}}
\subsection{Model representation}
This section reviews the univariate and multivariate mixed causal-noncausal models that have been considered in the literature.\\
\indent Univariate mixed causal and noncausal models, for a strictly stationary zero mean series $y_t$, for $t=1, \dots, T$, were introduced by \cite{breidt1991maximum} as the following autoregressive models:
\begin{equation}
y_{t}=\sum_{j=1}^{p}\theta_{j}y_{t-j}+\eta_{t},%
\end{equation}
where the autoregressive polynomial $\theta(z)=1-\sum_{j=1}^{p}\theta_{j}%
z^{j}$ has no roots on the unit circle ($\theta (z) \neq 0$ for $|z|=1$) and $\eta_t$ is $i.i.d.$, non-Gaussian. The process (1) is said to be purely causal (noncausal) if $\theta(z)$ has all the roots outside (inside) the unit circle. Process (1) is said to be mixed causal and noncausal when $\theta(z)$ has roots both inside and outside the unit circle. Because of the presence of the noncausal component, the right-hand side of equation (1) no longer represents the conditional expectation of the linear model. Consequently, this model becomes capable of capturing nonlinear dynamics, including local trends (bubbles) and conditional heteroscedasticity (see \cite{breidt1991maximum}, \cite{lanne2011noncausal}, \cite{hencic2015noncausal}, and \cite{gourieroux2018misspecification}).\\
\indent As shown in \cite{breid1991maximum}, the mixed causal and noncausal AR($p$) process defined in (1) admits a unique two-sided strictly stationary solution:
\begin{equation}
    y_t = \sum_{j=-\infty}^{\infty} \psi_j \eta_{t+j},
\end{equation}
with $\psi_0$ equal to 1. The two-sided MA representation clarifies why the autoregressive process (1) is mixed causal and noncausal: in addition to being affected by past and present shocks, the dependent variable is also affected by future shocks. In the event the process in (1) is purely causal (resp. noncausal), then $\psi_j = 0$ for all $j < 0$ (resp $j > 0$) (\cite{breidt1991maximum}).\\
\indent As \cite{findley1986uniqueness} pointed out, the coefficients of a two-sided moving average representation (including the present, past, and future errors) can be distinguished from a one-sided moving average representation (including the past and present errors) only if the error term $\eta_t$ follows a non-Gaussian distribution. This distinction arises because Gaussian distributions are entirely characterized by their second-order moments, which display symmetry over time in stationary processes. Therefore, in Gaussian processes, distinguishing between the backward and look-forward representations is not possible (see \cite{giancaterini2022climate}). As a consequence, any estimator relying solely on the linear second-order properties of the system, such as the OLS estimator, does not possess the capability to discern this feature. Hence, mixed causal and noncausal processes can always be represented as conventional purely causal AR($p$) (resp. purely noncausal) with the same linear autocovariance function as the true $DGP$. In addition, this conventional causal process (resp. noncausal) shares the characteristic of having the same roots as those outside the unit circle of the $DGP$, along with the inverses of the roots of the $DGP$ that lie inside the unit circle (resp. the same roots of the $DGP$ that lie inside the unit circle, along with the inverse of the roots of the $DGP$ that lie outside the unit circle).  In particular, this applies to all mixed causal and noncausal AR($p$) processes with autoregressive coefficients achieved by considering all combinations obtained by replacing the true roots with their reciprocals (see \cite{breidt1991maximum}). However, among all these processes sharing the same linear autocovariance functions as the true $DGP$, only the correct specification provides an $i.i.d.$ error term. This is why, in addition to the non-Gaussianity assumption, the correct identification of the noncausal component also requires an $i.i.d.$ error term (\cite{hecq2016identification}).\\
\indent Extending (1) to the multivariate framework results in a strictly stationary $n-$dimensional mixed causal and noncausal process expressed as:
\begin{equation}
    Y_t=\Theta_1 Y_{t-1}- \dots \Theta_p Y_{t-p}+u_t,
\end{equation} 
with det($\Theta (z)$) having roots inside ($n_1$) and outside ($n_2$) the unit circle. In addition, we assume the error term possesses a variance-covariance matrix denoted as $\Sigma_u$. It should be noted that also in the multivariate framework case, the assumptions of $i.i.d.$ and non-Gaussianity of the error term, $\{ u_t \}_{t=1}^{T}$, are required to correctly identify the noncausal component of the process or, equivalently, the roots inside the unit circle. Indeed, all the properties highlighted previously in the univariate framework can easily be extended to the multivariate framework (see \cite{davis2020noncausal}, \cite{gourieroux2017noncausal}).\\
\indent The existence of a stationary solution of (1)-(3), as well as the two-sided moving average representation of $Y$, is provided by \cite{gourieroux2017noncausal}. We reproduce their representation theorem to set up notations.\\\\

\textbf{Representation theorem} \textit{(\cite{gourieroux2017noncausal}}): When an $n$-$dimensional$ mixed causal and noncausal VAR(1) is considered (with $n \geq 1$), there exists an invertible $(n\times n)$ real matrix $A$, and two square real matrices: $J_{1}$ of dimension ($n_{1}\times n_{1}$) and $J_{2}$ of 
dimension ($n_{2}\times n_{2}$) such that all eigenvalues of $J_{1}$ (resp. $J_{2}$) are those of $\Theta_0$ with a modulus strictly less (resp. larger) than 1, and such that:
\begin{equation}
	Y_{t} =  A_{1}Y_{1,t}^{*}+ A_{2} Y_{2,t}^{*}
\end{equation}
\begin{equation}
	Y_{1,t}^{*} =  J_{1}Y_{1,t-1}^{*}+ u_{1,t}^{*}, \ \ \       Y_{2,t}^{*} =  J_{2}^{-1}Y_{2,t+1}^{*}-J_{2}^{-1} u_{2,t+1}^{*}
\end{equation}
\begin{equation}
	Y_{1,t}^{*} =  A^{1}Y_{t}, \ \ \ Y_{2,t}^{*} =  A^{2}Y_{t}
\end{equation}
 \begin{equation}
	u_{1,t}^{*} =  A^{1}u_{t}, \ \ \ Y_{2,t}^{*} =  A^{2}u_{t}
\end{equation}
where $[A_{1},A_{2}]=A$ and $[A^{1\prime},A^{2\prime}]^{\prime}=A^{-1}$. \\

\indent In Equation (5), it is evident that $Y_{1,t}^*$ and $Y_{2,t}^*$ represent the purely causal and purely noncausal components of the processes, respectively. Consequently, these two components can be interpreted as the causal and noncausal elements of the process 
$Y_t$. For the mixed causal and noncausal framework $VAR(p)$, with $p \geq 2$, the extension can be achieved by utilizing the companion form and rewriting the process as mixed causal and noncausal VAR(1) (\cite{gourieroux2017noncausal}).\\

\indent Note that mixed causal and noncausal processes have an alternative multiplicative representation (\cite{lanne2011noncausal}), \cite{hecq2016identification}, \cite{gourieroux2016filtering}, \cite{lanne2013noncausal}), \cite{nyberg2014forecasting}, and \cite{cubadda2023detecting}). In the univariate framework, the two representations overlap, allowing transitioning between them (\cite{hecq2022spectral}). However, in the multivariate framework, as emphasized by \cite{davis2020noncausal}, \cite{rygh2022causal}, \cite{cubadda2023detecting}, and \cite{gourieroux2022nonlinear}, the alternative multiplicative representation does not always exist and covers only a subset of mixed causal and noncausal processes. This makes the specification in equation (3) more general compared to the multiplicative representation. Therefore, this paper does not consider the alternative multiplicative representation.

\subsection{GCov estimator}
Section 2.1 emphasizes that the identification of mixed causal and noncausal processes using only linear second-order moments is not possible. However, it also points out that within processes sharing the same linear autocovariance function, only the true model has an $i.i.d.$ error term. While distinguishing the true model based on linear second-order moments is not feasible, examining the second moments and second cross-moments of nonlinear functions of the $i.i.d.$ non-Gaussian errors can identify the process (\cite{chan2006note}). This concept underlies the $GCov$ estimators introduced by \cite{gourieroux2017noncausal} and \cite{gourieroux2022generalized} and denoted as $GCov17$ and $GCov22$, respectively. The $GCov22$ estimator extends Portmanteau type tests (e.g. Ljung–Box) to a multivariate framework with non-linear transformations of the autocovariances. It minimizes the following Portmanteau statistic to estimate a $n-dimensional$ mixed causal and noncausal VAR($1$) process:
\begin{equation}
	\hat{\Theta}= \underset{\Theta}{\mathrm{argmin}} \sum_{h=1}^{H} Tr \bigl[ \hat{\Gamma}(h) \hat{\Gamma}(0)^{-1}\hat{\Gamma}(h)' \hat{\Gamma}(0)^{-1} \bigr],
\end{equation}
where, $H$ is the highest selected lag, $\hat{\Gamma}(h)$ is the sample autocovariance between $a(u_t)$ and $a(u_{t-h})$, with $a(u_t)= \bigl[ a_1(u_{t})^{\prime}, \dots , a_K(u_t)^{\prime}\bigr]$, and $a_j(u_t)$ is an element by element function, for $j=1, \dots , K$. Hence, $K$ indicates the number of linear and nonlinear transformations included in the $GCov$ estimator. $Tr$ computes the trace of a matrix. The choice of an informative set of transformations ($a_{j}$) depends on the specific series under investigation. As highlighted by \cite{gourieroux2017noncausal} and \cite{gourieroux2022generalized}, this problem is analogous to selecting moments in the Generalized Method of Moments (GMM) estimation or instruments in Instrumental Variable (IV) estimation. For instance, in financial applications aiming to capture the absence of a leverage effect, one can select both linear and quadratic functions. Specifically, assuming $n=2$, we may consider the following set of four functions ($K=4$): $a_1(u_t) = u_{1,t}$, $a_2(u_t) = u_{2,t}$, $a_3(u_t) = u_{1,t}^2$, and $a_4(u_t) = u_{2,t}^2$. This implies that $a_1$ and $a_2$ provide the errors of the two variables in the investigated process. Meanwhile, $a_3$ operates on the error term of the first variable, $u_{1,t}$, by squaring it for each $t=1, \dots, T$, where $T$ represents the total number of observations. Similarly, the function $a_4$ emulates the behavior of $a_3$, except that it applies the squaring operation to $u_{2,t}$ for each $t=1, \dots, T$. Alternatively, one can consider the signs of returns and their squares to separate volatility dynamics from the bid-ask bounce effect: \(a_1(u_t) = \text{sign}(u_1)\), \(a_2(u_t) = \text{sign}(u_2)\), \(a_3(u_t) = u_1^2\), and \(a_4(u_t) = u_2^2\). It is important to note that if $a(u_t)$ solely includes linear transformations of the error term, $\hat{\Gamma}(h)$, with $h=1, \dots, H$, would only provide information about the linear second-order moments of the process, rendering the estimator unable to identify and estimate the correct specification. Therefore, the introduction of nonlinear transformations serves the purpose of enabling us to estimate the true process. For the asymptotic normality of $GCov22$, finite fourth moments of $a(u_t)$ are required (\cite{gourieroux2022generalized}). In addition, the estimator in (8) is semi-parametrically efficient.\\
\indent The matrix in (8) is diagonalizable, with the sum of its eigenvalues being the sum of the squares of the canonical correlations between $a(u_{t})$ and $a(u_{t-h})$, for $h=1, \dots, H$. For comparison, the $GCov17$ estimator minimizes:
\begin{equation}
	\hat{\Theta}= \underset{\Theta}{\mathrm{argmin}} \sum_{h=1}^{H} Tr \bigl[  \hat{\Gamma} (h)  
     diag(\hat{\Gamma}(0))^{-1} 
  \hat{\Gamma}(h)^{\prime} 
     diag(\hat{\Gamma}(0))^{-1} 
 \Bigr],
\end{equation}
where $diag(\hat{\Gamma}(0))$ is the matrix with solely the diagonal elements of $\hat{\Gamma}(0)$. Hence, the only difference between (8) and (9) is that $Gcov17$ only takes into account the diagonal elements of matrix $\hat{\Gamma}(0)$. This feature makes $GCov17$ particularly appealing in the high-dimension framework, where a larger size of matrix $\hat{\Gamma}(0)$ is involved, potentially leading to numerically more stable computation. Furthermore, the estimator in (9) is consistent, and asymptotically normally distributed when the 4th-order moments of $a(u_t)$ are finite. In general estimator (9) is not semi-parametrically efficient, except when the weights in the objective function are $\hat{\Gamma}(0)$, instead of $diag \hat{\Gamma}(0)$ and the estimators in (8) and (9) coincide  (see \cite{gourieroux2017noncausal}). Note that for mixed causal and noncausal processes $VAR(p)$, with $p \geq 2$, we can use the companion matrix to rewrite the process as mixed causal and noncausal VAR($1$) so that the estimators (8) and (9) can easily be applied to the new representation of the model.\\
\indent It is important to recollect that our objective functions (8) and (9) are multivariate Portmanteau test statistics. A minimization of them will provide the estimated coefficient matrices with roots inside and outside the unit circle. We do not use (8) and (9) as usual misspecification tests after having estimated parameters by OLS for instance. Indeed second order estimates cannot find roots inside the unit circle. Finally, the $GCov$ estimator is an attractive alternative to the Continuously Updating Generalized Method of Moments (CUGMM) estimator (see \cite{hansen1996finite}).

\section{\textbf{Finite sample performances of Gcov22}}
We examine the behavior of the $GCov$ estimator for mixed causal and noncausal processes. Our investigation considers both multivariate and univariate frameworks.  As highlighted in Section 1, our focus in the univariate framework is solely to offer a clear visual depiction of the objective function in a two-dimensional Cartesian plane. For this purpose, we consider a purely noncausal AR($1$) process. In the multivariate framework, without loss of generality, we assume a variance-covariance matrix of the error term, $\Sigma_u$, to be diagonal and serially and cross-sectionally $i.i.d.$. This assumption is typically made when considering a structural VAR process.\\
\indent We have compared $Gcov22$ and $Gcov17$ and find that the estimators perform similarly in both the univariate and structural multivariate models. However, it is worth noting that $Gcov22$ exhibits better performance than $Gcov17$ in processes with a  non-diagonal covariance matrix. This superiority arises from $Gcov22$'s ability to consider $\hat{\Gamma}(0)$ as a whole, allowing it to capture dependencies among the error terms in a more flexible manner. In this paper, we primarily focus on the $GCov22$ estimator, but similar results, available upon request, are obtained with $GCov17$.

\subsection{The univariate framework}
Let us consider the following purely noncausal process, with $n=1$:
\begin{equation}
y_t = 1.5 y_{t-1} + \eta_t,
\end{equation}
where $\eta_t$ follows an $i.i.d.$ Generalized Student's-$t$ distribution with a scale parameter $\sigma=5$ and degrees of freedom $\nu =5$. Despite having one root inside the unit circle, namely $1.5^{-1}$, this process does not diverge to infinity due to the Representation theorem presented in Section 2. Specifically, by applying equations (4)-(5) to process (10), we find that the $DGP$ is given by: $y_t=(1.5)^{-1} y_{t+1}-(1.5)^{-1}\eta_{t+1}$. This expression is characterized by one root outside the unit circle, thereby yielding a stationary solution (see also \cite{lanne2011noncausal}).\\
\indent Next, we employ a grid search strategy for the autoregressive coefficient, spanning a range from -1 to 5 with a step size of 0.01. Note that even if the autoregressive coefficient in reverse time equals $1.5$ is larger than 1, the process does not explode to infinity because of the correlation between $y_{t-1}$ and the error term. For each coefficient within this interval, we calculate the objective function value. This computation is conducted for $T=500$ observations and using $H=10$ in equation (8). 
This procedure enables us to visually represent the objective function of the $GCov$ estimator and analyze the potential existence of local minima.\\
\indent In empirical investigations, we do not have prior knowledge regarding the most suitable functions $a_j$ and the appropriate value of $K$ for our dataset. Indeed, as mentioned in Section 2, the choice depends on the specific process under investigation. Therefore, we explore various combinations of $a_k$ and $K$:
\begin{itemize}
\item T0:   $a_{1}(\eta_t)$=$\eta_t$;

\item T1: $a_{1}(\eta_t)$=$\eta_t$,
	   $a_{2}(\eta_t)$=$\eta_t^{2}$,
	   $a_{3}(\eta_t)$=$\eta_t^{3}$,
	   $a_{4}(\eta_t)$=$\eta_t^{4}$;
 
 \item T2:  $a_{1}(\eta_t)$=$\eta_t$,
            $a_{2}(\eta_t)$=$log(\eta_t^{2})$;

 \item T3:  $a_{1}(\eta_t)$=$sign(\eta_{t})$,
            $a_{2}(\eta_t)$=$\eta_t^{2}$;
 
 \item T4: $a_{1}(\eta_t)$=$sign(\eta_t)$,
	$a_{2}(\eta_t)$=$log(\eta_t^{2})$.
\end{itemize}
Figure 1-(a) confirms that when only linear transformations of the error term (T0) are employed, the $GCov$ is unable to differentiate between the true noncausal process and the conventional causal autoregressive process with coefficient $1.5^{-1}$. This is equivalent to compare values that maximise the likelihood function in OLS regressions on respectively only leads or lags. We would obtain estimates around $1.5^{-1}$ in both time directions.
Then, the objective function of $GCov22$ exhibits two global minima. However, even when both linear and nonlinear transformations (T1-T4) are included, the bimodality issue in the $GCov22$ objective function is not completely resolved. In this scenario, Figure 1-(a) illustrates that the objective function has a local minimum at $1.5^{-1}$
and a global minimum of 1.5. This pattern remains true, irrespective of the specific choices of $a_k$ and $K$ (similar results were obtained for various $a_k$ and $K$ beyond T0-T4, not presented here but available upon request). Whether the problem with the local minimum is related to the choice of $a_j$, $K$, or a combination of both, it is crucial to recognize that in empirical investigations where the selection of these inputs is not known in advance, we may encounter challenges with local minima, resulting in inaccurate estimates.\\
\indent In Figure 1-(b), we present the empirical density function of the estimated parameter obtained through a Monte Carlo simulation with $N=1000$ replications, assuming $p$ as known. In each replication, the initial guess for the optimization algorithm is derived from the OLS estimate of the process, denoted as $\theta_{OLS}$. As emphasized in Section 2, the OLS estimate, which considers only the linear second-order moments of the process, falls short in identifying the noncausal component, thereby providing an initial guess close to the local minimum. Once $\theta_{OLS}$ is obtained, we use it as the initial value for the $GCCov$ estimator, which is optimized using the BFGS optimization algorithm. The BFGS algorithm is a well-established deterministic optimization technique that approximates the inverse gradient of the objective function to locate the minimum. It initiates with an initial minimum estimate and iteratively refines this estimate using gradient information and an approximation of the inverse Hessian matrix. We also optimize the $GCov$ with other common numerical optimization algorithms, such as the Nelder-Mead, conjugate gradient method, and limited-memory BFGS. However, as they yielded similar results, those findings are available upon request. Figure 1-(b) and Table 1 highlight the difficulty in the convergence of the optimization algorithm towards the global minimum.\\
\indent The interpretation of the results suggests that local minimum issues arise due to the $GCov$ estimator's domain being divided into two sets within this specific $DGP$; set 1 with coefficients $\theta< 1$ and set 2 with coefficients $\theta > 1$.\\
\indent Each set contains a value of $\theta$ that minimizes the objective function when considering linear transformations of the autocovariance function (T0), that is $1.5^{-1}$ and $1.5$. However, when nonlinear autocovariance functions of the error term are considered (T1-T4) in (8), only $\theta=1.5$ represents the global minimum, while its causal counterpart acts as the local minimum. As a consequence, if the optimization problem starts in set 1, that is where a local minimum occurs, it is likely that the numerical optimization algorithm will get trapped in that set and converge to the local minimum instead of the global one.\\
\indent We conclude that a careful selection of an appropriate initial value and optimization algorithm is essential in guaranteeing accurate parameter estimates. This is even more important in the multivariate analysis, as we will show in Section 3.2. Indeed, the inclusion of a larger number of parameters in multivariate analysis adds a layer of complexity, making the task of achieving convergence towards the global minimum a more intricate challenge.
\begin{figure}[H]
    \centering
    \caption{\textbf{Application of $\boldsymbol{Gcov22}$ to a purely noncausal AR(1))}}
    \begin{subfigure}{0.38\linewidth}  
        \includegraphics[width=\linewidth]{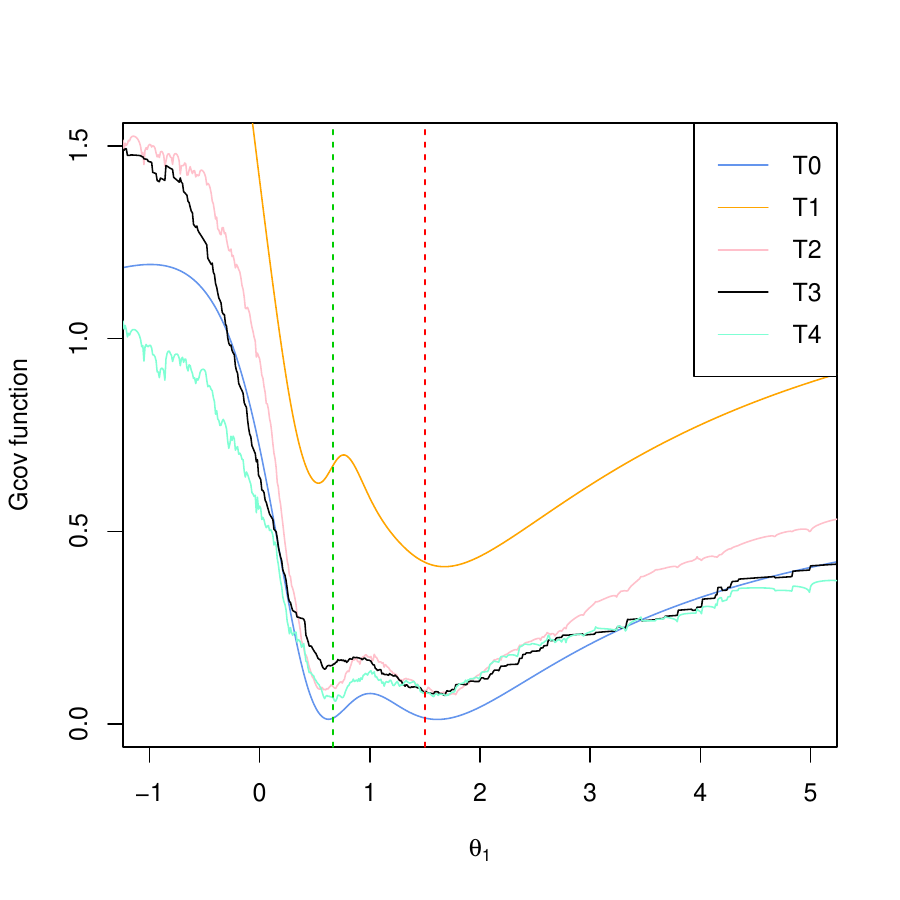}
        \caption{Visualization of the $Gcov22$\\ 
        objective function.\\
        The red and green vertical lines \\ 
        represent the true and inverse \\values
        of $\theta$, respectively.}
    \end{subfigure} \ \ 
    \begin{subfigure}{0.38\linewidth}  
        \includegraphics[width=\linewidth]{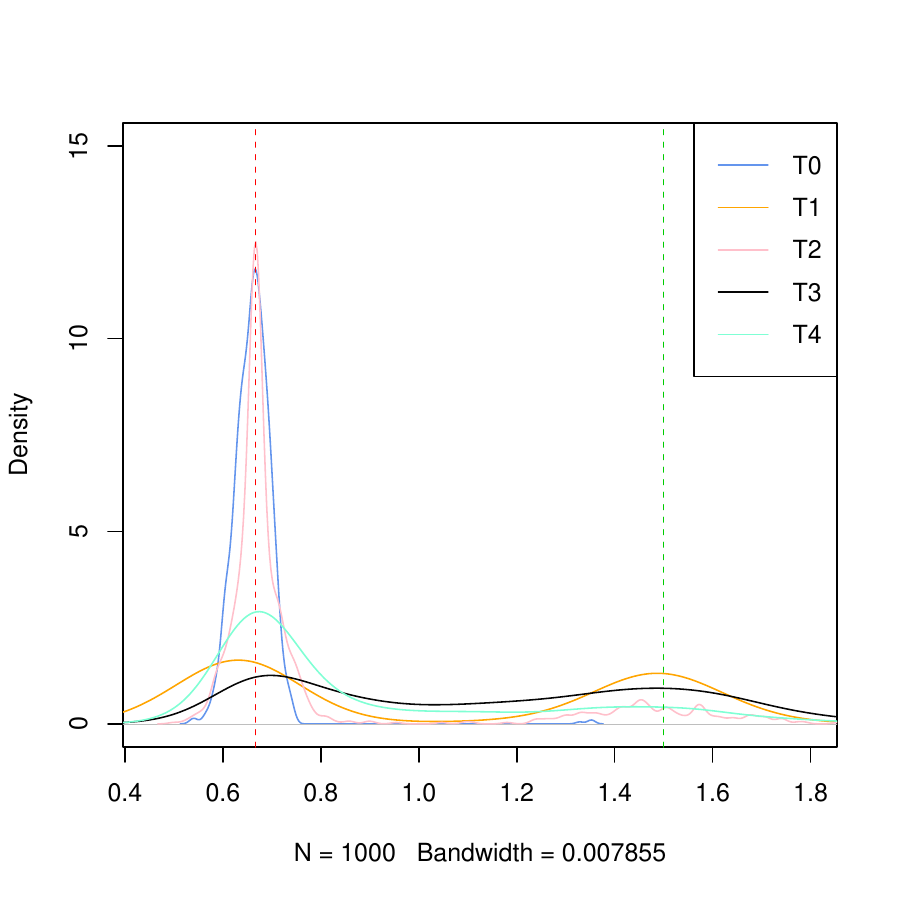}
        \caption{Empirical Density of $\hat{\theta}$\\: A Monte Carlo \\experiment with $N=1000$\\ replications and\\ $T=500$ observations.}
    \end{subfigure}
\end{figure}

\subsection{The multivariate framework}
We next investigate the performance of the $GCov22$ estimator for multivariate models. We consider a $3-dimensional$ mixed causal and noncausal  VAR($1$):
\begin{equation}
    Y_t=\Theta Y_{t-1}+u_t,
\end{equation}
where $u_t$ is serially $i.i.d.$, distributed according to a multivariate Student's-$t$ distribution. The error distribution is characterized by $\nu=4$ degrees of freedom, a scale parameter equal to $\sigma=2$, and a diagonal variance-covariance matrix $\Sigma_u$. The sample size is $T=500$ and $\Theta$
\begin{equation}
    \Theta=\begin{bmatrix}
        \theta_{11} & \theta_{12} & \theta_{13}\\
        \theta_{21} & \theta_{22} & \theta_{23}\\
        \theta_{31} &\theta_{32} & \theta_{33}
    \end{bmatrix} =
    \begin{bmatrix}
        3.97 & -3.73 & 1.3\\
        2.29 & -2.38 & 1.41\\
        1.87 &-2.16 & 1.40
    \end{bmatrix} , 
\end{equation}
with two eigenvalues inside the unit circle (related to the causal component): $j_1 = 0.3$ and $j_2 = 0.5$, and one eigenvalue outside the unit circle (related to the non-causal component): $j_3 = 2.2$. Appendix A displays the graph and autocorrelation function of one simulation of a process $Y_t$.\\
\indent In the multivariate framework, it is not possible to graphically represent the objective function of the $GCov22$ estimator as a function of $\Theta$ since the domain is now in the space $\mathbb{R}^{3 \times 3}$. However, we calculate the empirical density function of the matrix $\Theta$. In this context, we compute the density function of $\hat{\Theta}$ by implementing a Monte Carlo experiment with $N=1000$ replications. We use the OLS estimate of $\Theta$ ($\Theta_{OLS}$) as the initial guess for the optimization of the $GCov$ estimator, specifically the BFGS optimization algorithm. Figure 2 and Table 1 display the results when considering nonlinear transformations T1-T4. We are not interested in the linear transformation T0 since, as seen in the previous sections, it is not able to capture the true $DGP$ in mixed causal and noncausal processes. It should be noted that in this case, the number of $K$ increases since we implement the transformation to the error term of each series. As a consequence, T1 is now characterized by $K=12$, while all the other transformations are by $K=6$. The results show that similarly to the univariate framework, the $GCov22$ objective function has local minima when the true roots (or true eigenvalues) of the matrices are replaced by their reciprocals. More specifically, Figure 2 illustrates that the empirical density function of $\Theta$ is centered around the initial guess $\Theta_{OLS}$, characterized by eigenvalues $j_1$, $j_2$, and $j_3^{-1}$, rather than the population matrix $\Theta$ expressed in (12). Table 1 summarizes these findings, and shows that the process is predominantly erroneously identified as purely causal by $GCov22$. However, as in Section 3.1, the transformation $T1$ performs better in identifying the true process (Table 1), although the density function of $\hat{\Theta}$ still displays its peaks mainly centered around $\Theta_{OLS}$ instead that the true population matrix.\\
\indent For comparison, Figure 3 displays the density function of $\hat{\Theta}$ when the noncausal counterpart of the matrix in (12) is selected as the initial point for the optimization of the $GCov22$. Specifically, to obtain this matrix, it is necessary to reverse the process and estimate it as a VAR($1$) by OLS, and then invert the estimated matrix. We denote this matrix as $\tilde{\Theta}$, which is characterized by eigenvalues outside the unit circle: $j_{1}^{-1}$, $j_{2}^{-1}$, and $j_{3}$. Even in this case, the empirical density function has been obtained by the use of an MC experiment, setting $N=1000$ replications. Using $\tilde{\Theta}$ as the initial guess makes us erroneously identify the process as purely noncausal most of the time, and obtain an empirical density function developed around $\tilde{\Theta}$, instead of the matrix in (12).\\
\indent In Figure 4, the depicted empirical density function is obtained by using the population matrix as the initial value. The results highlight that in this scenario, the conventional optimization algorithm converges successfully, yielding an empirical density function that closely aligns with (12). As a result, the model is correctly identified most of the time, regardless of the nonlinear transformations T1-T4 employed (see Table 1).\\
\indent The results extend the findings of the univariate framework to the multivariate one: the objective function domain of the $GCov22$ estimator consists of four sets, characterized by the matrices yielding specific roots:
\begin{itemize}
\item Set 1: Characterized by all those autoregressive matrices that provide eigenvalues inside the unit circle (resp. roots outside the unit circle);
\item Set 2: Characterized by all those autoregressive matrices that provide two eigenvalues inside the unit circle and one eigenvalue inside the unit circle;
\item Set 3: Characterized by all those autoregressive matrices that provide one eigenvalue inside the unit circle and two eigenvalues inside the unit circle;
\item Set 4: Characterized by all those autoregressive matrices that provide eigenvalues outside the unit circle.
\end{itemize}
Each set contains a matrix that minimizes the value of the objective function when considering linear transformations of the autocovariance. However, when nonlinear transformations of the autocovariance of the error term are considered in (8), only (12) represents the global minimum, while the others act as local minima. Hence, in our case, if the optimization algorithm initiates within a set where a local minimum is located (Sets 1-3-4), conventional optimization algorithms are likely to become trapped in that set and converge to the local minimum instead of the global one. Conversely, successful convergence is achieved when the initial value is selected from the same set as the global minimum (Set 4).  Therefore, the choice of the starting point for the optimization algorithm and the optimization algorithm itself are two crucial steps to avoid identification issues, potentially preventing the misidentification and misestimation of the investigated process.\\
\indent 
\begin{figure}[H]
    \centering
    \caption{\textbf{Density function of $\hat{\Theta}$: initial guess $\boldsymbol{\Theta}_{OLS}$}}
    \includegraphics[width=\linewidth]{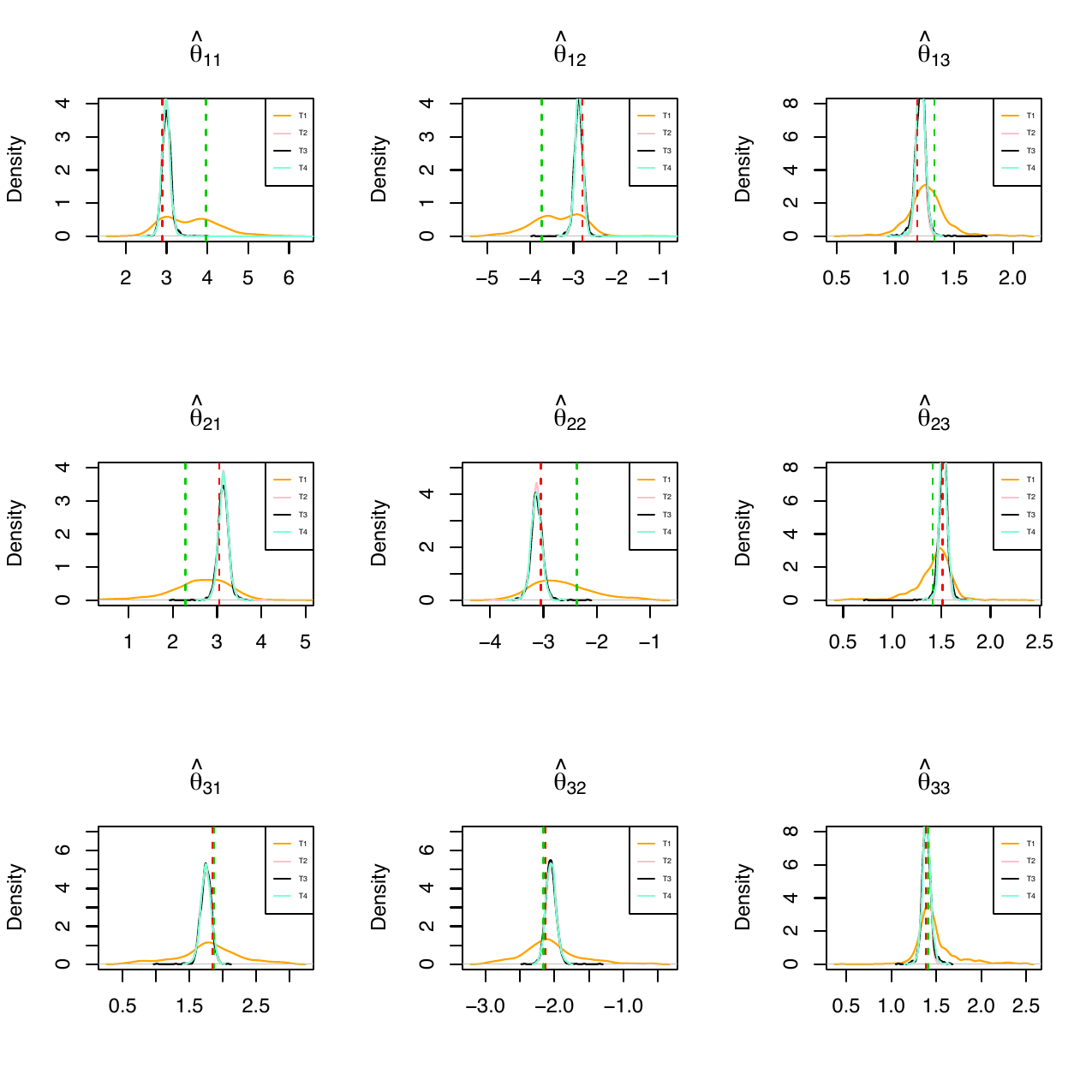}
    \caption*{\textit{The empirical density function of $\Theta$ is depicted, considering the population matrix as defined in (12), marked with vertical green dashed lines. The initial value for the BFGS optimization algorithm is set as the causal counterpart of the matrix (12), i.e., $\Theta_{OLS}$, shown with red dashed lines. The analysis is based on a sample size of $T=500$ observations.}}
\end{figure}
\begin{figure}[H]
    \centering
    \caption{\textbf{Density function of $\hat{\Theta}$: initial value $\boldsymbol{\tilde{\Theta}}$}}
    \includegraphics[width=\linewidth]{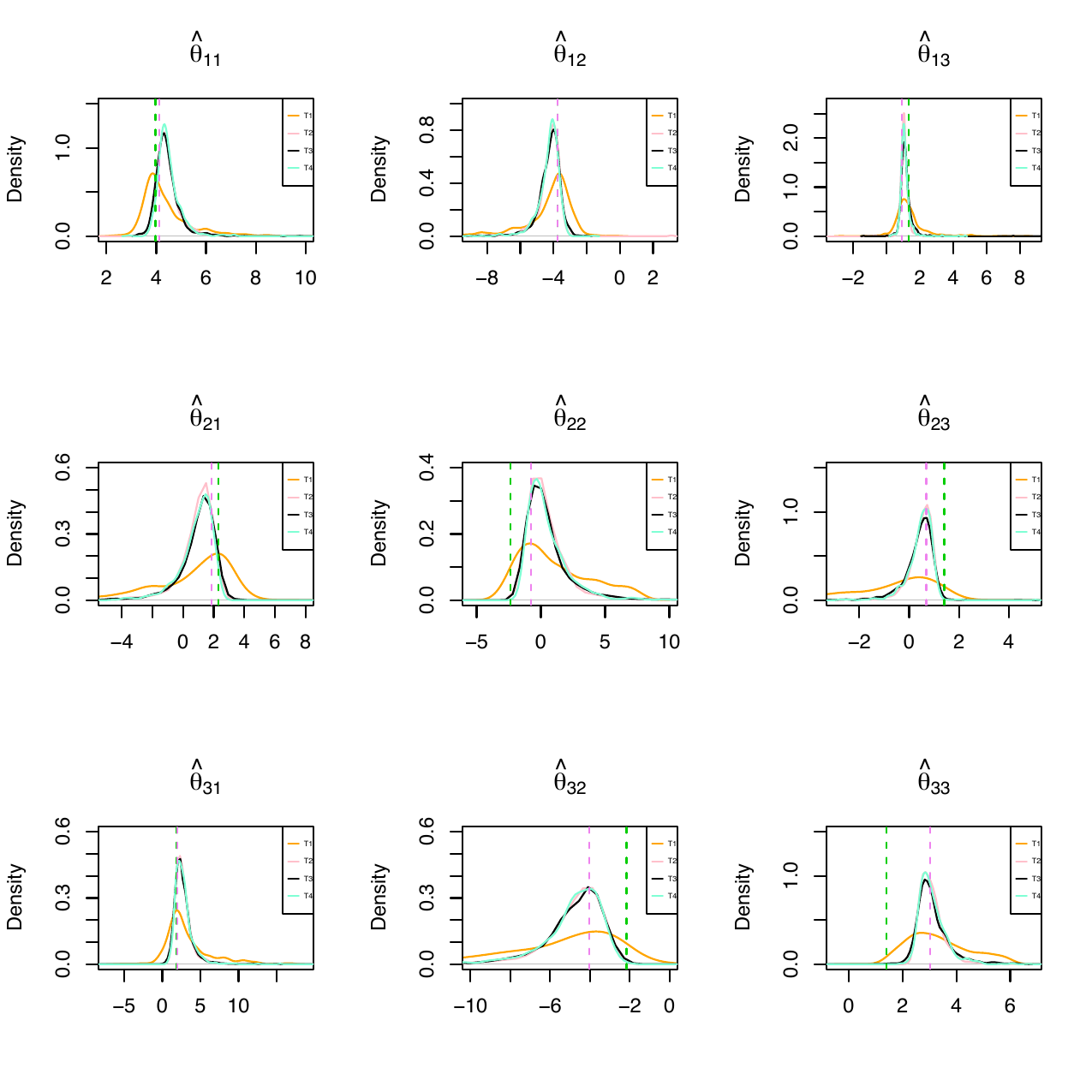}
    \caption*{\textit{The empirical density function of $\Theta$ is depicted, considering the population matrix as defined in (12), marked with vertical green dashed lines. The initial value for the BFGS optimization algorithm is set as the noncausal counterpart of the matrix (12), $\tilde{\Theta}$, shown with violet dashed lines. The analysis is based on a sample size of $T=500$ observations.}}
\end{figure}
\begin{figure}[H]
    \centering
    \caption{\textbf{Density function of $\hat{\Theta}$: initial value $\boldsymbol{\Theta}$ as in (12)}}
    \includegraphics[width=\linewidth]{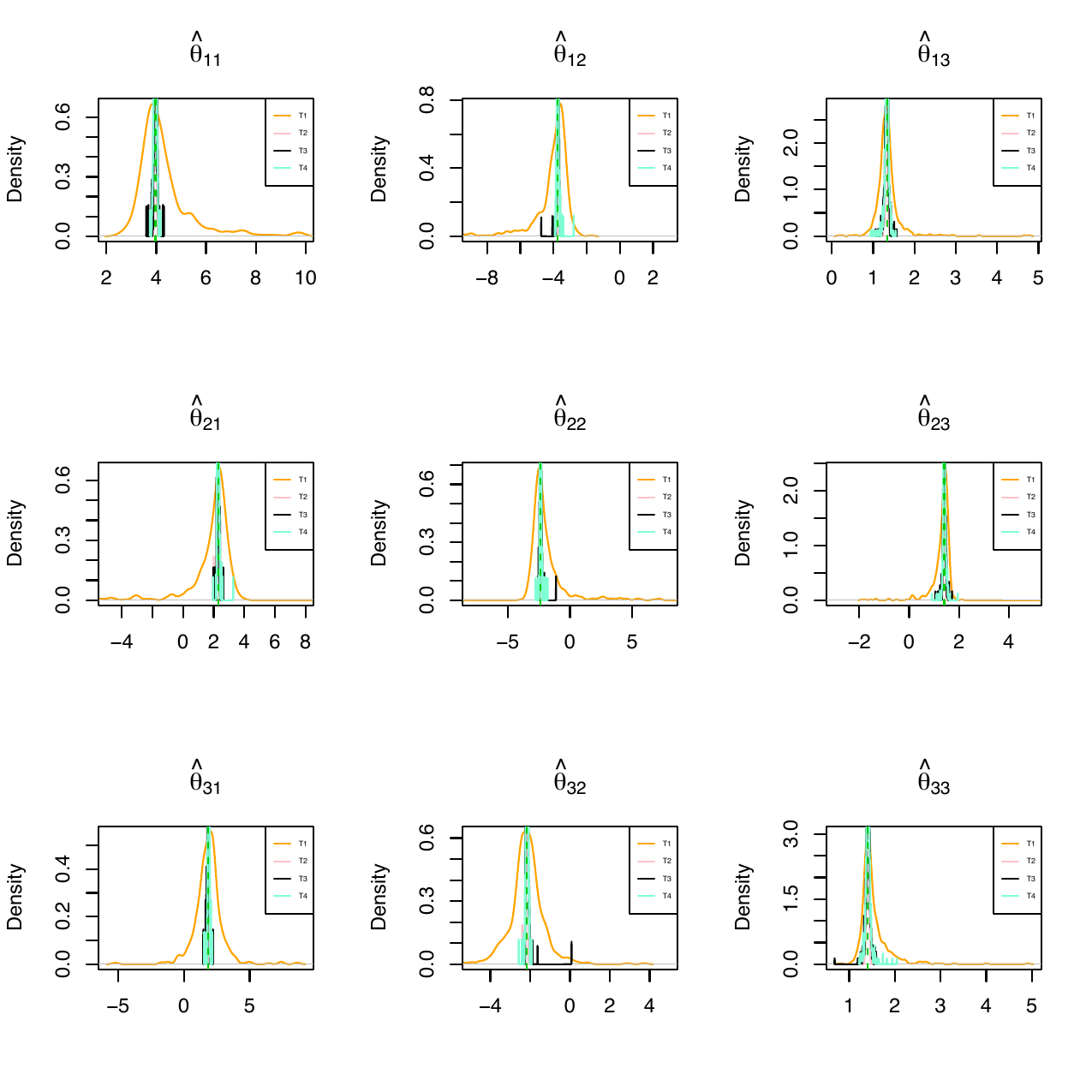}
    \caption*{\textit{The empirical density function of $\Theta$ is depicted, based on the population matrix defined in (12), marked with vertical green dashed lines. Here, the matrix in (12) serves a dual purpose as both the population matrix and the initial point for the optimization algorithm.}}
\end{figure}

\section{\textbf{Simulated Annealing}}
In the previous section, we stressed the importance of selecting an initial point that belongs to the same set as the global minimum. Specifically, selecting an initial guess of a matrix with the same $n_1$ and $n_2$ as the population matrix is crucial for achieving successful convergence of the BFGS optimization algorithm. However, in empirical investigations, determining the number of roots that lie within and outside the unit circle of the population matrix at the prior can be challenging. Therefore, in this section, we investigate the performance of the SA optimization algorithm (\cite{kirkpatrick1983optimization}, \cite{vcerny1985thermodynamical}, and \cite{goffe1994global}) when applied to the $GCov$ estimator. Our particular focus will be on the nonlinear transformation T1, which, as demonstrated in Section 3 outperforms T2-T4, in addition to being the most commonly employed nonlinear transformation (see \cite{gourieroux2022generalized} and \cite{gourieroux2022nonlinear}).

\subsection{The alghoritm}
\indent SA is an optimization method that was inspired by the annealing process used in metallurgy. In metallurgy, materials are cooled gradually to eliminate any imperfections and achieve a more stable state. The algorithm starts at a high temperature ($T^o$) and gradually cools down over time to reduce the likelihood of getting stuck in a local minimum. Hence, in optimization problems, $T^o$ is a parameter that controls the search space exploration during optimization. When $T^o$ is high, the algorithm is more likely to accept worse solutions than the current one, enabling it to escape local optima and explore new areas of the search space. As $T^o$ decreases, the algorithm is less likely to accept suboptimal solutions and converge toward the global optimum. However, if the cooling rate is too high, the algorithm may not be able to escape local minima (see \cite{corana1987minimizing}, \cite{goffe1992simulated}, \cite{goffe1994global}, and \cite{goffe1996simann}).\\
\indent Let us now explain how the SA algorithm works when applied to the $GCov$ estimator. We consider a mixed causal and noncausal VAR($1$) as (11), and we use $f$ to denote the objective function of either $GCov17$ or $GCov22$. Furthermore, we denote the maximum and minimum temperature values of $T^{o}$ as $T^{o}_{MAX}$ and $T^{o}_{MIN}$, respectively.\\
\indent To initiate the optimization process, a function evaluation is performed at the randomly selected starting point, denoted as $\Theta^{S}$. Subsequently, a new matrix $\Theta$ ($\Theta^{\prime}$) is computed. Specifically, it is determined by adjusting the $ij$-$th$ element of the matrix $\Theta^{\prime}$ ($\theta^{\prime}_{ij}$) using the following equation:
\begin{equation}
\theta^{\prime}_{ij} = \theta^{S}_{ij} + m_{ij} \ \ \ \forall i, j = 1, \dots, n.
\end{equation}
Here, $\theta^{S}_{ij}$ is the $ij$-$th$ element of matrix $\Theta_S$, and $m_{ij}$ is randomly selected from a uniform distribution within the interval $[m_{MIN}, m_{MAX]}$. The value $f(\Theta^{\prime})$ is then calculated and compared with $f(\Theta^{S})$. If $f(\Theta^{\prime}) < f(\Theta^{S})$, $\Theta^{\prime}$ is accepted, and the algorithm progresses downhill. In the opposite scenario, when $f(\Theta^{\prime}) > f(\Theta^{S})$, the potential acceptance of $\Theta^{\prime}$ is determined using the Metropolis criterion. According to this criterion, we compute the variable $p^o$ as follows:
\begin{equation}
    p^o = e^{-\frac{(f(\Theta^{\prime}) - f(\Theta^S))}{T^{o}}},
\end{equation}
we then compare it with $p^{*}$, that is, a number randomly selected from the range [0, 1]. If $p^o < p^{*}$, $\Theta^{\prime}$ is rejected, and the algorithm remains at the current point in the function. Conversely, if $p^o > p^{*}$, we accept $\Theta^{\prime}$ and move downhill. Equation (14) illustrates why a lower value of $T^o$ decreases the probability of making an uphill move. To find the optimal solution, the procedure is repeated $M$ times for each $T^{o}$, starting from $T^{o}_{MAX}$ and gradually reducing it at a rate of $r$, for a total of $Q$ times, until it reaches $T^{o}_{MIN}$.\\
\indent SA unlike conventional optimization algorithms has the potential to let us escape from local minima (see \cite{corana1987minimizing}, \cite{aarts2005simulated}). However, as a drawback, the parameters associated with the SA method, such as $\theta_{MIN}$, $\theta_{MAX}$, $T^{o}_{MAX}$, $r$, $Q,$ and $M$, are typically treated as black-box functions and are contingent upon the objective function to be minimized. In empirical studies, a common approach to investigate whether the global minimum has been found is to repeat the algorithm with a different initial state $\Theta^{S}$. If the same global minimum is reached, it can be concluded with high confidence that convergence has been achieved. In cases where a different result is obtained, it may be necessary to modify one or more of the parameters involved in the SA algorithm.\\

\subsection{Performance of SA: univariate framework}
In this section, we evaluate the performance of the SA algorithm minimizing the $GCov22$ objective function for univariate mixed causal and noncausal models. To this end, we conduct a Monte Carlo experiment to calculate the empirical density functions of $\hat{\theta}$, while maintaining the same purely noncausal model as specified in (10). The coefficient $\theta$ obtained by SA referred to as $\theta_{SA}$, serves as the starting point for the BFGS optimization of the the $GCov22$ estimator. This strategy offers two distinct benefits. Firstly, it provides an opportunity to explore the impact of different initial value strategies on reaching the global optimum, thereby facilitating comparison with the results presented in Figure 1 and Table. Secondly, it allows for the refinement of the solution obtained through the SA method. This refinement proves particularly valuable when $\theta_{SA}$ is in proximity to the global minimum, but there remains room for improvement in its solution.\\
\indent  As previously mentioned, we begin with an initial temperature of $T^{o}_{MAX}$, and in each of the $Q$ iterations, we allow it to decrease at a rate of $r$. After $Q$ iterations, it reaches the minimum temperature, denoted as $T^{o}_{MIN}$. It is worth noting that, in our approach, the final temperature $T^{o}_{MIN}$ is a deterministic function of $T^{o}_{MAX}$, $r$, and $Q$. This is true because $T^{o}_{MIN}$ is obtained through $Q$ reductions at a rate of $r$ from the initial value $T^{o}_{MAX}$, i.e., $T^{o}_{MIN} = T^{o}_{MIN}(T^{o}_{MAX}, r, Q)$.\\
\indent In the literature, it is common to employ a cooling rate of $r=0.85$, as indicated in \cite{goffe1994global} and \cite{corana1987minimizing}. However, determining suitable values for $T^{o}_{MAX}$ and $Q$, which subsequently determine $T^{o}_{MIN}$, often requires an empirical approach. Therefore, before conducting our MC experiment, we undertake a preliminary analysis of the behavior of the $DGP$ under investigation by initially setting $T^{o}_{MAX}$ and $Q$ to high values. This allows us to monitor the performance of the objective function throughout the optimization process. More specifically, in this preliminary analysis, we set $T^{o}_{MAX}=5000$ and $Q=150$. Figure 5-(a) illustrates the behavior of the objective function of $GCov22$ as a function of the number of iterations $Q$. The objective value of $GCov22$, from the first approximately 50 iterations, fluctuates around an average value of 2.7 without leading to any improvements for our optimization problem. This suggests that $T^{o}_{MAX}=5000$ is excessively high and results in inefficient use of time. Around $Q=50$, corresponding to $T^{o}=1.5$, the objective function decreases and converges toward the global optimum. These results are further confirmed by Figure 5-(b), which depicts the behavior of the parameter $\theta$ as a function of $Q$. Based on the insights gained from this analysis, we set $T_{MAX}=1.5$ and $Q=100$ in our MC experiment. Furthermore, to effectively explore the search space we set $M=100$. The choice of a high value for $M$ is crucial for a comprehensive exploration of the search space. The results are presented in Figure 5-(c) and summarized in Table 1. Now, the SA algorithm yields a significant improvement in results compared to Section 3.1: $\hat{\theta}$ closely approximates the true value, and the true dynamic is captured $90\%$ of the time. It should be noted that cases where the $GCov22$ incorrectly identifies our process as purely causal may arise from certain replications of our MC experiment where the objective function necessitates higher values of either $T^{o}_{MAX}$, $Q$, $M$, or a combination of them. As previously mentioned, these parameters are typically problem-specific, and their selection involves experimentation. Nevertheless, for practical reasons, we maintain the same values for $Q$ and $M$ across all replications.

\subsection{Performance of SA: multivariate framework}
In this section, we evaluate the performance of SA applied to $GCov22$ estimation of multivariate mixed causal and noncausal models. As in Section 3.2, in each Monte Carlo replication, we simulate our time series and then estimate it using GCov, with SA serving as the initial value for the BFGS optimization problem of the $GCov22$(see Section 3.1 for the reasons of such a combination of SA and BFGS optimization algorithms). For the sake of comparison, we maintain the same Monte Carlo inputs and autoregressive model as specified in Section 3.2. In this way, we can explore the impact of different initial value choices on finding the global optimum, in comparison with the results presented in Figures 2-3-4 and Table 1.\\
\indent Using the approach employed for the univariate framework, we set $T_{MAX}=800$, $Q=200$, and $M=2000$. The results are displayed in Figure 6 and summarized in Table 1. Even in this case, the SA algorithm yields a significant improvement in results compared to Section 3.1. Indeed, the density of the estimator is now centered around the population value (12). Lastly, as in the previous section, it is worth noting that, for practical reasons, we maintain a constant value for $T^{o}_{MAX}$, $Q$, and $M$ in each replication of the Monte Carlo experiment.
\begin{figure}[H]
    \centering
    \caption{\textbf{Performance of SA when implemented to an univariate noncausal process}}
    \begin{subfigure}{0.38\linewidth}  
        \includegraphics[width=\linewidth]{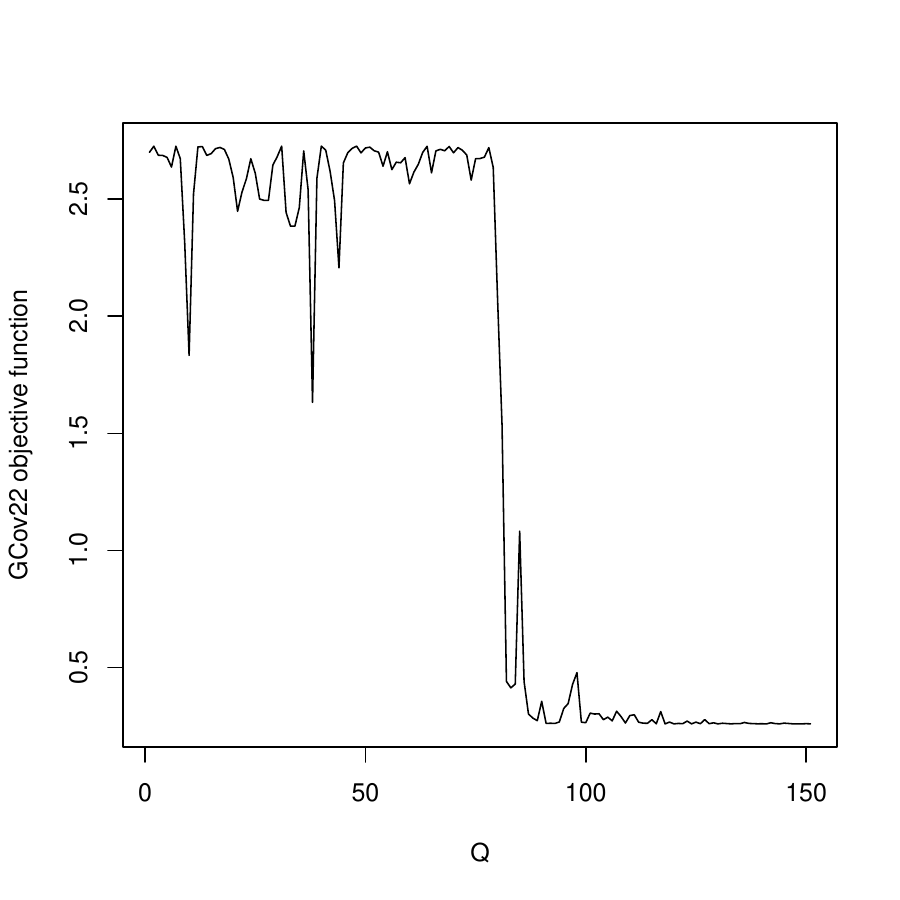}
        \caption{\textit{Visualization of the $Gcov22$\\                  objective when expressed as a \\ 
            function of $Q$,
                during the SA\\ optimization algorithm.}}
    \end{subfigure}
    \begin{subfigure}{0.38\linewidth}  
        \includegraphics[width=\linewidth]{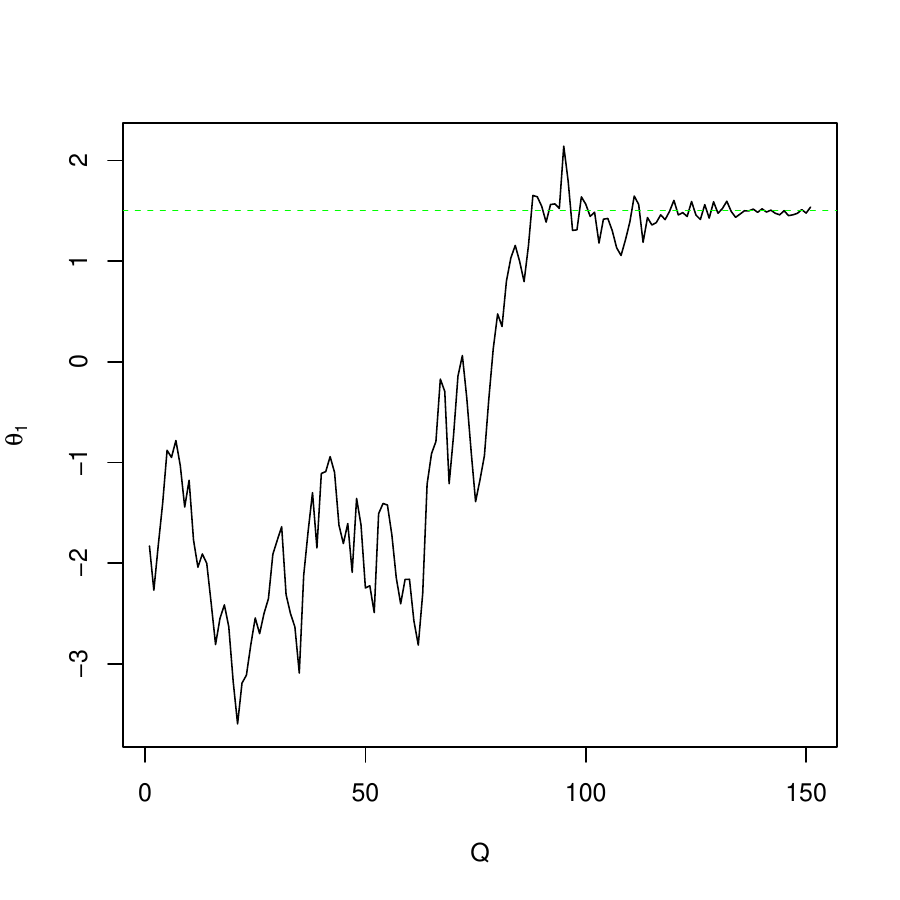}
       \caption{\textit{Visualization of the $\theta$ when \\                   expressed as a function of $Q$,\\ during the SA optimization\\ algorithm.}}
    \end{subfigure}
    \begin{subfigure}{0.38\linewidth}  
        \includegraphics[width=\linewidth]{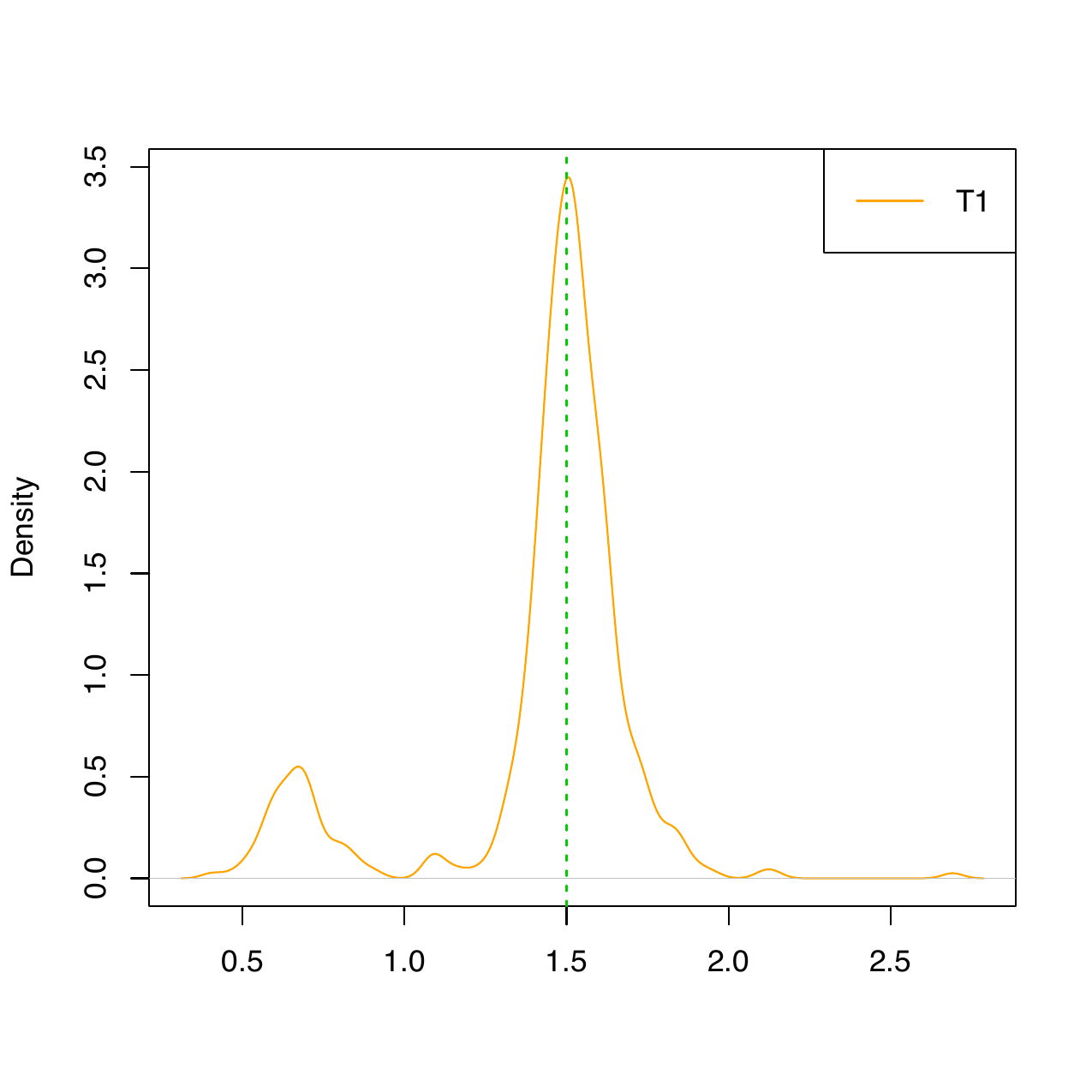}
        \caption{\textit{Empirical Density of $\hat{\theta}$ when SA is \\
                employed to capture the initial guess. The vertical
                dashed line indicates the true
                value $\theta=1.5$}}
    \end{subfigure}
\end{figure}

\begin{figure}[H]
	\begin{center}
        \caption{Empirical density function of $\Theta$ with $SA$}
		\includegraphics[width=\linewidth]{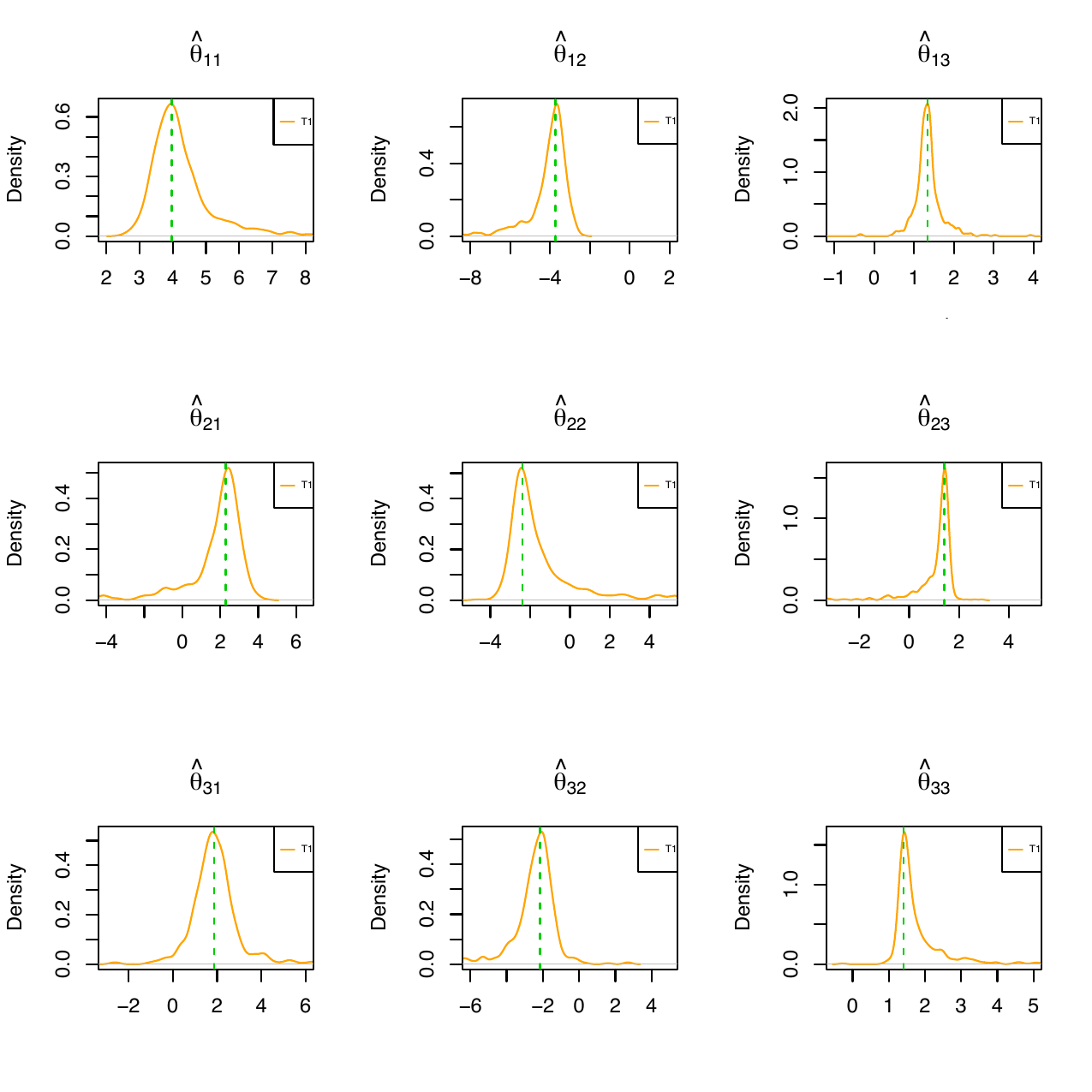}
	\end{center}
\caption*{\textit{The empirical density function of $\hat{\Theta}$ is obtained using $Gcov22$ when the SA strategy is implemented to select the starting values. We assume that $\Theta$, as defined in (12), represents the population matrix. The vertical green lines in the plot indicate the corresponding population values.}}

\end{figure}
\begin{table}[H]
\caption{\textbf{Estimated dynamics}}
\resizebox{\textwidth}{!}{%
\begin{tabular}{cccccc}
\hline
\multicolumn{6}{c}{Univariate framework: purely noncausal AR(1)} \\ \hline
Initial guess    & $a_k$ & \multicolumn{2}{c}{Purely causal AR($1$)} & \multicolumn{2}{c}{Purely noncausal AR(1)} \\ \hline
$\theta_{OLS}$   & T0    & \multicolumn{2}{c}{$100\%$}   & \multicolumn{2}{c}{$0\%$}     \\
$\theta_{OLS}$   & T1    & \multicolumn{2}{c}{$52.8\%$}  & \multicolumn{2}{c}{$47.2\%$}  \\
$\theta_{OLS}$   & T2    & \multicolumn{2}{c}{$89.1\%$}  & \multicolumn{2}{c}{$10.1\%$}  \\
$\theta_{OLS}$   & T3    & \multicolumn{2}{c}{$44.3\%$}  & \multicolumn{2}{c}{$55.7\%$}  \\
$\theta_{OLS}$   & T4    & \multicolumn{2}{c}{$79.0\%$}  & \multicolumn{2}{c}{$21.0\%$}  \\ \hline
$\theta$     & T1    & \multicolumn{2}{c}{$0\%$}     & \multicolumn{2}{c}{$100\%$}   \\ \hline
$\theta_{SA}$    & T1    & \multicolumn{2}{c}{$12.0\%$}  & \multicolumn{2}{c}{$88.0\%$}  \\ \hline
\multicolumn{6}{c}{Multivariate framework: VAR($n_1=2,n_2=1,p=1$)}                                   \\ \hline
Initial guess    & $a_k$ & VAR($n_1=3,n_2=0,p=1$)  & VAR($n_1=2,n_2=1,p=1$)  & VAR($n_1=1,n_2=2,p=1$)  & VAR($n_1=0,n_2=3,p=1$)  \\ \hline
$\Theta_{OLS}$   & T1    & $34.7\%$      & $55\%$        & $7.6\%$       & $0.7$         \\
$\Theta_{OLS}$   & T2    & $100\%$       & $0.0\%$       & $0.0\%$       & $0.0\%$       \\
$\Theta_{OLS}$   & T3    & $95.5\%$      & $4.3\%$       & $0.2\%$       & $0.0\%$       \\
$\Theta_{OLS}$   & T4    & $96.7\%$      & $3.2\%$       & $0.0\%$       & $0.1\%$       \\ \hline
$\tilde{\Theta}$ & T1    & $0.1\%$       & $2.3\%$       & $48.6\%$      & $49\%$        \\
$\tilde{\Theta}$ & T2    & $0.0\%$       & $0.0\%$       & $0.0\%$       & $100\%$       \\
$\tilde{\Theta}$ & T3    & $0.0\%$       & $0.0\%$       & $7.0\%$       & $93\%$        \\
$\tilde{\Theta}$ & T4    & $0.0\%$       & $0.0\%$       & $0.5\%$       & $99.5\%$      \\ \hline
$\Theta$       & T1    & $0.0\%$       & $94.5\%$      & $3.1\%$       & $2.4\%$       \\
$\Theta$       & T2    & $0.0\%$       & $97.5\%$      & $2.4\%$       & $0.1\%$       \\
$\Theta$       & T3    & $0.0\%$       & $97.6\%$      & $2.3\%$       & $0.1\%$       \\
$\Theta$       & T4    & $0.0\%$       & $97.3\%$      & $2.3\%$       & $0.4\%$       \\ \hline
$\Theta_{SA}$    & T1    & $1.4\%$       & $64.6\%$      & $28.4\%$      & $5.6$         \\ \hline
\end{tabular}}
\caption*{\textit{The table illustrates the performance of $GCov22$ in estimating the mixed causal and noncausal processes. The column initial guess indicates the strategy adopted to capture the initial point for the optimization algorithm, while $a_k$ indicates the linear and nonlinear transformations employed. In this table, VAR($n_1,n_2,1$) indicates a VAR(1), with $n_1$ roots outside the unit circle and $n_2$ roots inside the unit circle.}}
\end{table}

\section{\textbf{Empirical investigations}}
We conduct an empirical analysis on a bivariate time series consisting of 363 daily data points for wheat and soybean futures in US Dollars. The data spans from October 18, 2016, to March 29, 2018. The dataset was sourced from https://ca.finance.yahoo.com, with wheat futures represented by the ticker ZW=F and soybean futures by the ticker ZS=F. Figure 7 displays the demeaned data, while Figure 8 presents kernel-smoothed density estimators of the series.\\
\indent Our primary objective in this investigation is to evaluate the performance of the $GCov22$ estimator. In addition, we seek to identify the presence of speculative bubbles in agricultural commodity markets. Detecting such bubbles has significant implications for various stakeholders, including market participants, policymakers, and investors, as it directly impacts decision-making in both the agricultural and financial sectors. It is noteworthy that the examined series does not exhibit global trends or other widespread and persistent explosive patterns. Instead, they display localized trends and spikes, often sharing similar patterns with concurrent spikes. To gain insights into the interactions among these variables and determine whether noncausal components drive these processes, we proceed to estimate a $2-dimensional$ VAR($p$).\\
\indent Having identified $p=2$ as the appropriate order, we apply the $GCov22$ method to the demeaned data and reject the null hypothesis of the Gaussianity of the residuals. In this specific case, we use both the OLS estimate ($\Theta_{OLS}$) and the results obtained by the SA method as starting points for the BFGS optimization problems. The results are presented in Table 2. Notably, Table 2 highlights that the choice of the initial guess significantly affects the results. When we employ the OLS strategy to determine the initial guess, the $GCov22$ method identifies the process as purely causal VAR($2$). However, when we utilize the SA method, we obtain a lower value of the $GCov22$ objective function, and the bivariate process is identified as a mixed causal and noncausal VAR($2$) with three roots outside the unit circle and one root inside the unit circle. Indeed, the eigenvalues in the latter case lie both inside and outside the unit circle: $j_1=0.972$, $j_2=0.88$, $j_3=0.604$, and $j_4=-4.355$. These findings underscore the importance of combining SA with $GCov22$. Employing SA to capture the initial guess is crucial in this case, enabling us to identify a non-causal component of the process, which, in turn, allows us to capture the nonlinear features that define these series.
\begin{figure}[H]
\centering
    \caption{\textbf{Empirical investigation: wheat and soybean}}
        \includegraphics[width=7cm]{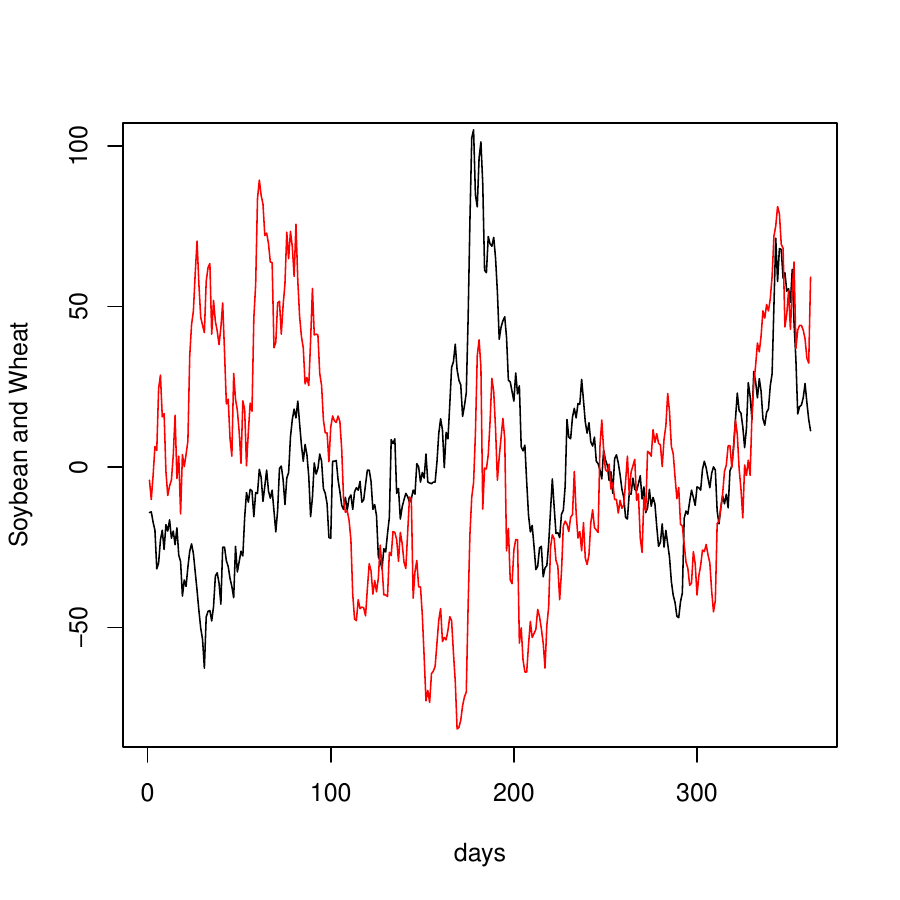}
        \caption*{\textit{The graph shows the demeaned prices of wheat (black line) and soybean (red line) futures from October 18, 2016, to March 29, 2018.}}
\end{figure}
\begin{figure}[H]
\centering
 \caption{\textbf{Marginal sample densities of demeaned daily future price series.}}
        \includegraphics[width=7cm]{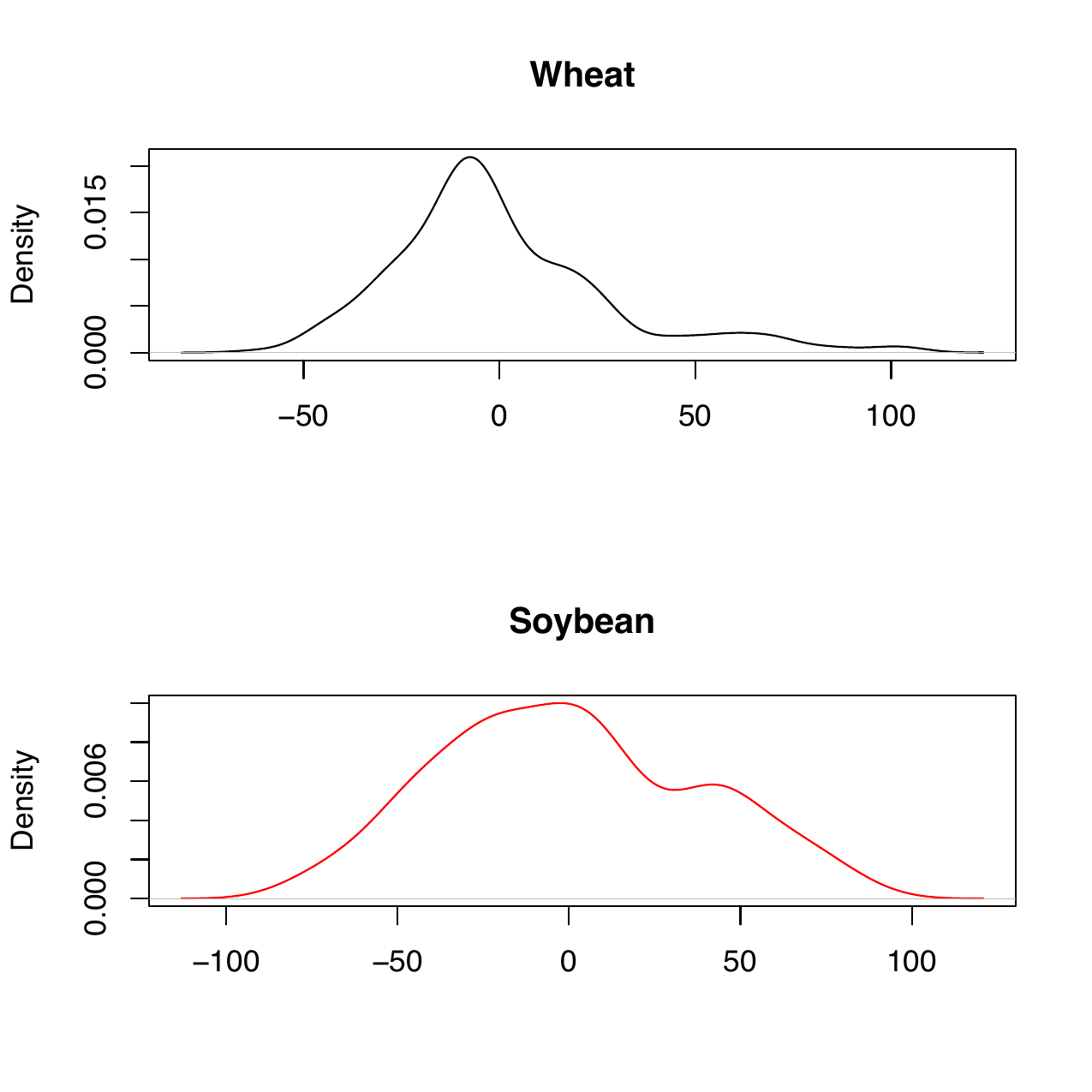}
        \caption{Marginal sample densities of demeaned daily future price series.}
\end{figure}

\begin{table}[H]
\center
\caption{\textbf{Estimated coefficients of our empirical analysis}}
\begin{tabular}{lllllllll}
\hline
\multicolumn{9}{c}{Soybean and Wheat}     
\\ \hline
SV                                       &                & \multicolumn{2}{c}{$\hat\Phi_{1}$} & \multicolumn{2}{c}{$\hat\Phi_{2}$} &                  &  & \\ \hline

                                     &                & $\phi_{j,1}$    & $\phi_{j,2}$   & $\phi_{j,1}$                       & $\phi_{j,2}$ &&  f.v.& Model       \\ \hline
$\Theta_{OLS}$                     & $\phi_{1,j}$ & 1.16                & 0.24               & -0.29                                   & -0.29             &  & 2.00 & VAR($n_1=4,n_2=0,p=2$)    \\
                                         & $\phi_{2,j}$ & 0.06                & 1.05               & -0.04                                   & -0.09             &  &    &       \\ \hline
SA                    & $\phi_{1,j}$ & 0.44                & 1.23               & 0.52                                   & -1.21             &     &1.50 &     VAR($n_1=3,n_2=1,p=2$)  \\
                                         & $\phi_{2,j}$ & 3.31                & -2.34               & -3.19                                   & 3.10             &  &    &       \\ \hline
\end{tabular}%
\caption*{\textit{The column} ”SV” \textit{and} ”f.v.” \textit{display the strategy adopted to capture the starting values and the value of the function at the estimated values. In this table, VAR($n_1,n_2,1$) indicates a VAR(1), with $n_1$ roots outside the unit circle and $n_2$ roots inside the unit circle.}}
\end{table}

\section{\textbf{Conclusions}}
In this paper, we have investigated the performance of the $GCov$ in estimating mixed causal and noncausal models. The $GCov$ estimator, being a semi-parametric method, offers the advantage of not assuming any specific error distribution. Utilizing a portmanteau-type criterion based on nonlinear autocovariances ensures consistent estimates and consequently allows for the identification of the causal and noncausal orders of the mixed VAR.\\
\indent Our findings highlight the importance of considering an adequate number and type of nonlinear autocovariances in the objective function of the $GCov$ estimator. When these autocovariances are insufficient or inadequate, or when the error density closely resembles the Gaussian distribution, identification issues can arise. This manifests in the presence of local minima in the objective function, occurring at parameter values associated with incorrect causal and noncausal orders. Consequently, the optimization algorithm may converge to a local minimum, leading to inaccurate estimates.\\
\indent To overcome the problem of local minima and improve the estimation accuracy of mixed VAR models, we propose the use of the SA optimization algorithm as an alternative to conventional numerical optimization methods. The SA algorithm effectively manages the identification issues caused by local minima, successfully eliminating their effects. By exploring the parameter space more robustly and flexibly, SA provides a reliable solution for obtaining more accurate estimates of the causal and noncausal orders.\\
\indent The paper applies the $GCov22$ estimator to a bivariate commodity price series. The results highlight the existence of local minima in the bivariate processes, emphasizing the role of the SA algorithm in providing reliable results in empirical research.
\newpage
\section*{Appendix A}
\begin{figure}[H]
	\begin{center}
        \caption{\textbf{Graph of the simulated time series in (11)-(12)}}
		\includegraphics[width=0.5\linewidth]{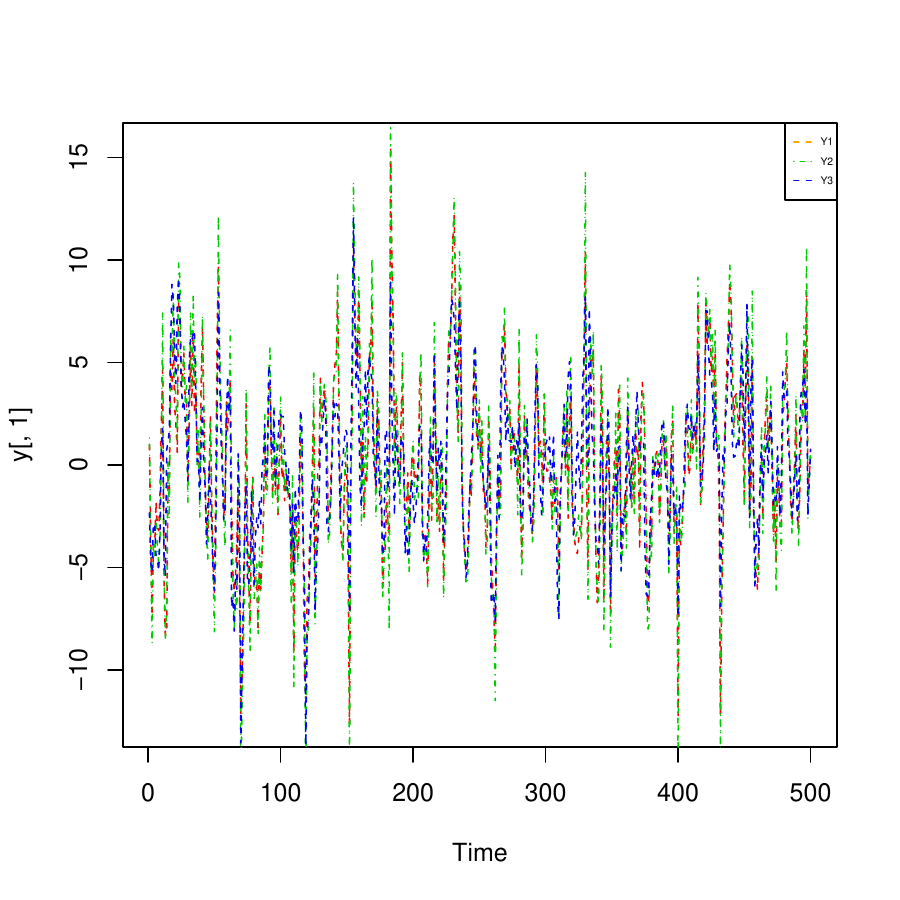}
	\end{center}
\end{figure}

\begin{figure}[H]
	\begin{center}
        \caption{\textbf{Autocorrelation function of process generated by (11)-(12)}}
		\includegraphics[width=0.5\linewidth]{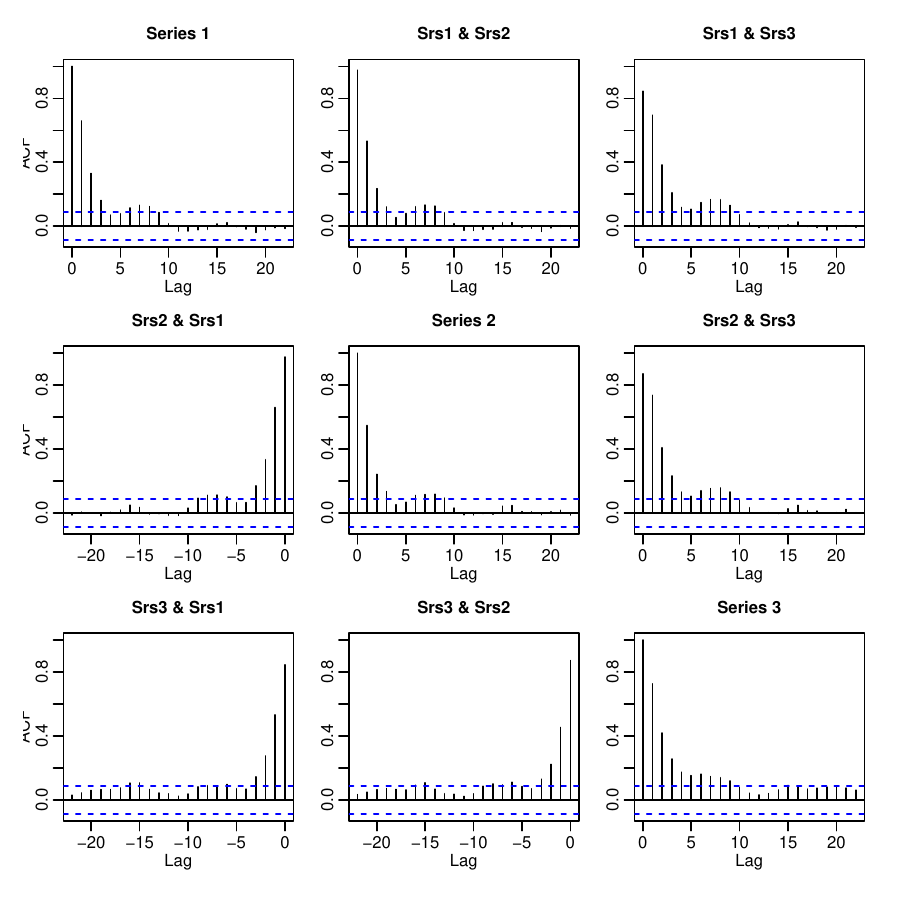}
	\end{center}
\end{figure}
\newpage
\biboptions{authoryear}
\raggedright
\bibliography{MAIN_NEW}

\end{document}